\documentstyle[twoside,sprocl]{article} 

\begin{document} 

\title{COSMOLOGICALLY SCREWY LIGHT} 

\author{BORGE NODLAND}

\address{Department of Physics and Astronomy, and Rochester Theory
Center for Optical Science and Engineering, University of Rochester,
Rochester, NY 14627}

\author{JOHN P. RALSTON}

\address{Department of Physics and Astronomy, and Kansas Institute for
Theoretical and Computational Science, University of Kansas, Lawrence,
KS 66044}

\maketitle 

\abstracts{
We present evidence for a new phenomenon in the propagation of
electromagnetic radiation across the Universe, a corkscrew rotation of
the plane of polarization not accounted for by conventional physics.}

\abstracts{[Published in Proceedings of the 14th Particles and Nuclei
International Conference (PANIC), Williamsburg, Virginia, May 22-28,
1996. Eds.: C. Carlson et al., World Scientific, Singapore (1997)]}

Electromagnetic radiation propagating on cosmological distances
provides an exceedingly sensitive laboratory for new phenomena. Here
we summarize observation of an unexpected systematic rotation of the
plane of polarization of radio signals from galaxies with redshift $z
>0.3.$  We find that the data are explainable in two ways: either invoking
unnatural conspiracies among the sources, or proposing new physics.

Radio polarization is produced by synchrotron radiation. However, the
observed plane of polarization from distant sources does not usually
align with the symmetry axis of the source (denoted $\psi$).  For
decades this has been studied in terms of Faraday rotation in the
intervening medium.  We study a residual quantity remaining after the
Faraday effect is taken out.

Extracting Faraday rotation does not depend on models, because the
Faraday angle of rotation for wavelength $\lambda$ goes like $\alpha
\lambda^2$.  Consistent linear dependence on $\lambda^2$ is indeed
observed. The problem is that the data fit requires more: for each
source (i), the fits are given by $\theta_i(\lambda) =  \alpha_i
\lambda^2 + \chi_i.$  The residual polarization  angle $\chi_i$ does
not generally align with the galaxy major axis; statistics on the
differences $\chi_i-\psi_i\pm\pi$ have puzzled astronomers for 30
years \cite{gw}.

Analysis of the data is challenging, due to several complications. The
data set is highly non-uniform in its angular distribution on the sky
and in the distances $r(z)$ to the sources.  We employed Monte Carlo
methods to search for correlations \cite{nr}.  We make thousands
of fake data sets with random polarization and major axis orientations,
while keeping the positions of the galaxies the same as the real data.
We then calculate the probabilities of linear correlations observed in
the data relative to the random sets.

We find \cite{nr} an amount of residual rotation beta described by a
dipole rule $\beta = \frac{1}{2} (r/\Lambda_s) \cos (\gamma)$, where
$\gamma$ is the angle between the propagating wavevector and an axis
$\vec s$ fit to the data. The dependence on the propagation direction
is quite novel, indicating anisotropy. Making a cut on $z>0.3$, which
selects the most-distant half of the data set, we find a striking
correlation with probabilities that the observed correlation would be
produced by random angular fluctuations less than $10^{-3}$; the effect
is $3.7\sigma$.

A separate study eliminates bias from fitting the $s$-direction to the
data, by fitting the best $\vec s$-direction data-set-by-data-set in
the Monte Carlo, and calculating the probability of finding the data's
correlation relative to the optimized sets so constructed.  This gives
a probability less than 0.006, corresponding to $2.7 \sigma$.  

The fits to the parameters are $\Lambda_s=1.1 \times 10^{25} \;
(h_0/h)\; m$, where $\frac{h_0}{h}$ is the ratio of 100 $km (s^{-1}) \;
(Mpc)^{-1}$ to the Hubble constant, while $\vec s=(0^\circ \pm
20^\circ$ declination, $21\pm 2$ hrs right ascension). The scale
$\Lambda_s$ is 1/10 the Universe's size; the effect is far too small to
affect laboratory measurements.  We do not find a significant
correlation for $z<0.3$; in our full data set we also do not find a
significant correlation of $\beta = (const)\ \ r$.

We are not the first to find puzzling statistics in the residual angle
\cite{gw}; the problem is that our correlation cannot be reconciled
with conventional ideas.  Explaining the effect via population
dependence among the sources requires an unnatural conspiracy across
the sky.  The role of systematic errors, with the possibility that the
angular correlation might arise in a local effect, is crucial.  The
observers deny this possibility; more importantly, the proposal is
ruled out by the fact that the near half of the data does not have the
correlation.  New physics  explanations are available \cite{nr};
perhaps the most conventional proposal would be domain walls of an
axion--like condensate.  The prediction of exactly the dipole rule in a
different model \cite{nr}, and the search for a limit, led to the
present investigation.

\section*{Acknowledgments}

This research was supported by DOE Grant Number DE-FG02-85ER40214
and the Kansas Institute for Theoretical and Computational Science.

\end{document}